\documentclass[12pt]{iopart}

\usepackage[dvips]{graphicx}
\begin{document}
\title{Spin nutation effects in molecular nanomagnet$-$superconductor tunnel junctions}
\author{J. Abouie$^{1,2*}$, B. Abdollahipour$^{3}$\footnote{These authors have contributed equally to this work.}
and A. A. Rostami$^{4}$ }
\address{$^1$ Department of Physics,
Institute for Advanced Studies in Basic Sciences (IASBS), Zanjan 45137-66731, Iran\\
$^2$ School of physics, Institute for Research in Fundamental
Sciences (IPM), Tehran 19395-5531, Iran\\
$^3$ Faculty of Physics, University of Tabriz, Tabriz 51666-16471, Iran\\
$^4$ Department of Physics, Shahrood University of Technology,
Shahrood 36199-95161, Iran}
\begin{abstract}
We study the spin nutation effects of the molecular nanomagnet on
the Josephson current through a superconductor$|$molecular
nanomagnet$|$superconductor tunnel junction. We explicitly
demonstrate that due to the spin nutation of the molecular
nanomagnet two oscillatory terms emerge in the $ac$ Josephson
current in addition to the conventional $ac$ Josephson current. Some
resonances occur in the junction due to the interactions of the
transported quasiparticles with the bias voltage and molecular
nanomagnet spin dynamics. The appearance of them indicate that the
energy exchanged during these interactions is in the range of the
superconducting energy gap. We also show that the spin nutation is
able to convert the $ac$ Josephson current to a $dc$ one which is
interesting for applications.
\end{abstract}
\maketitle

\section{Introduction}

Molecular nanomagnets have recently attracted intense attentions due
to their applications in quantum information processing
\cite{Leuenberger01} and molecular spintronics \cite{Bogani08} which
is an emerging field to combine the facilities of molecular
junctions and spintronics \cite{Rocha05}. In the molecular
spintronics magnetic molecules are employed to manipulate the spin
degrees of freedom for exploiting various properties of the
molecular junctions. The long magnetization relaxation time of
molecular nanomagnets \cite{Chri00,Gatt07} and the small sizes of
the junctions consisting of them emerge interesting phenomena such
as, negative differential conductance and complete current
suppression \cite{Jo06,Heersche06} which are crucial in high-density
information storage \cite{Mannini09}, quantum computing
\cite{Ardavan07} and nanoelectronics.

One of the powerful methods to probe intrinsic properties of
molecular nanomagnets is the transport measurements through the
junctions consisting of them \cite{Henderson07,Haque11,Zyazin12}. An
example is the direct observation of their magnetic states and their
easy axis orientations by three-terminal measurements of charge
transport \cite{Burzuri12}. The electronic transport through
magnetic molecules connected to metallic leads have been
investigated by extensive theoretical works. The different aspects
of the molecular nanomagnets such as the effect of the exchange
coupling between spins of conduction electrons and the spin of the
molecular nanomagnets \cite{Kim04} have been studied as well as the
Coulomb blockade in a transport through a molecular nanomagnet
weakly coupled to a magnetic and a nonmagnetic lead \cite{Elste06},
the possibility of writing, storing, and reading spin information in
memory devices \cite{Timm06}, the Kondo effect in transport through
a single molecular nanomagnet strongly coupled to two metallic
electrodes \cite{Romeike06}, effects of spin Berry phase on the
electron tunneling \cite{Gonzalez07}, magnetic switching of the
molecular nanomagnets' spin by a spin-polarized current
\cite{Misiorny07}, the tunneling magnetoresistance \cite{Misiorny09}
and the effect on the transport of a soft vibrating mode of the
molecule \cite{Cornaglia07}. The current through the precessing
molecular nanomagnet connected to the metallic contacts has been
obtained by considering the molecular magnet as a classical spin
\cite{Zhu02}. It was shown that the spin precession causes to
modulate the conductance with two frequencies $\omega_L$ and
$2\omega_L$, where $\omega_L$ is a Larmor precession frequency. It
has been recognized that the electron-spin-resonance scanning
tunneling microscopy (ESR-STM) technique is capable of detecting the
precession of a single spin through the modulation of the tunneling
current \cite{Balatsky02}. The mechanisms underlying the ESR-STM
technique are the spin-orbit coupling and exchange interactions
between the localized spin and conduction electrons.

The coupling of the molecular nanomagnets to the superconducting
leads changes the transport properties of them via the Andreev
reflection at the contacts. The Josephson current through a
molecular nanomagnet connected to the superconducting leads has been
investigated experimentally in Refs. \cite{Kasumov05,Winkelmann09}.
Theoretical investigations of the Josephson current through an
isotropic molecular nanomagnet revealed an asymmetric phase diagram
in the exchange coupling \cite{Lee08}. The coupling of the spin
dynamics of molecular nanomagnet and the Josephson current can
reveal many interesting phenomena. It was found that the Josephson
current through the junctions consisting of two superconductors
having equal spin-triplet pairing symmetry is modulated by the spin
precession, whereas when both the leads are spin-singlet
superconductors the Josephson current remains unmodulated
\cite{Zhu03}. It was also shown that a circularly polarized {\it ac}
spin current with the Larmor precession frequency is generated in
the spin-singlet superconducting leads due to the spin precession
\cite{Teber10}. The dynamics of the Andreev bound states and their
effects on the current flowing through the junction have also been
studied in Refs. \cite{Holmqvist11,Holmqvist12,Stadler13}.

In this paper, we investigate the Josephson current through the
junction consisting of a molecular nanomagnet connected to two
spin-singlet superconductors via tunnel barriers. We present a
theoretical study of the molecule's spin nutation effects on the
flowing supercurrent. The spin nutation can be generated by applying
an external time dependent magnetic field to the molecular
nanomagnet which is a combination of a static and a rotating
transverse {\it rf} fields. Very recently, we have shown that the
spin nutation of a molecular nanomagnet could pump the charge
current through a Josephson junction \cite{AAA12}. Defining an
anomalous Green's function we obtain the Josephson current through
the tunnel junction at the presence of a bias voltage. We show that
the $ac$ Josephson current strongly depends on the spin nutation. We
explicitly demonstrate that due to the molecular nanomagnet spin
nutation two oscillatory terms are emerged in the $ac$ Josephson
current in addition to the conventional $ac$ Josephson oscillation.

The outline of this paper are as follows. In section \ref{model} we
introduce our model. We write Hamiltonian of the Josephson junction
and introduce the dynamics of the molecular nanomagnet. In the
section \ref{current} we obtain the $ac$ Josephson current. Results
and discussions are presented in section \ref{results}. Finally, we
summarize our results and give the conclusions.


\section{Model and basic equations}\label{model}

\begin{figure}[h]
\centerline{\includegraphics[width=7cm]{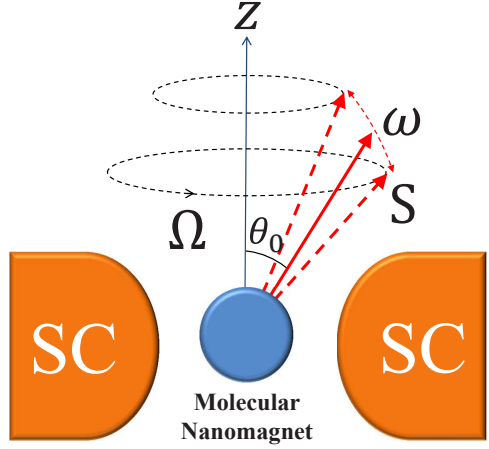}}
\caption{Schematic representation of the spin nutation of a
molecular nanomagnet coupled to the spin-singlet superconductors via
tunnel barriers.} \label{system-model}
\end{figure}

The system under consideration is a Josephson tunnel junction with a
nutating molecular nanomagnet sandwiched between two conventional
spin-singlet superconductors as shown in Fig.(\ref{system-model}).
The system is generally described by the Hamiltonian
\begin{equation}\label{total-hamiltonian}
H(t)=H_L+H_R+H_T(t)\,
\end{equation}
where $H_L$ and $H_R$ are the BCS Hamiltonian of left (L) and right
(R) superconductors with the amplitude of the pair potential
$\Delta$ and phases $\chi_L$ and $\chi_R$:
\begin{equation}\label{hamiltonian-leads}
H_{\alpha}=\sum_{k,\sigma=\uparrow,\downarrow}\varepsilon_k
c_{\alpha,k,\sigma}^{\dagger}c_{\alpha,k,\sigma}+\sum_k\left(\Delta_{\alpha}
c_{\alpha,k,\uparrow}^{\dagger}c_{\alpha,-k,\downarrow}^{\dagger}+h.c.\right)
\end{equation}
where $c_{\alpha,k,\sigma}^{\dagger}(c_{\alpha,k,\sigma})$ is the
creation (annihilation) operator of an electron in the lead
$\alpha=L, R$ with momentum $k$ and spin $\sigma$,  $\varepsilon_k$
is the energy of single conduction electron. The two leads are
weakly coupled via the tunneling Hamiltonian:
\begin{equation}\label{tunneling-hamiltonian}
H_T(t)=\sum_{k,k',\sigma\sigma'}\left(c_{R,k,\sigma}^{\dagger}
T_{\sigma\sigma'}(t)c_{L,k',\sigma'}+h.c.\right),
\end{equation}
%
%
%
where $T_{\sigma\sigma'}(t)$ is a component of the time dependent
tunneling matrix which transfers electrons through the system. When
a local spin is embedded into the tunneling barriers the tunneling
matrix can be written as
\begin{equation}\label{tunneling-matrix}
\hat{T}(t)=T_0\hat{\mathbf{1}}+T_S\hat{\mathbf{S}}(t)\cdot\hat{\mbox{\boldmath
$\sigma$}},
\end{equation}
where $\hat{\mathbf{1}}$ is a $2\times 2$ unit matrix,
$\hat{\mathbf{S}}(t)=\frac{\mathbf{S}}{|\mathbf{S}|}$ is the unit
vector along the molecular nanomagnet's spin and
$\hat{\mbox{\boldmath $\sigma$}}=(\sigma_x,\sigma_y,\sigma_z)$ is
the Pauli spin operator. The time dependence in
Eq.(\ref{tunneling-matrix}) originates from the nutational motion of
the molecular nanomagnet's spin localized in the junction. The spin
nutational motion which has been shown in Fig.(\ref{system-model})
is a combination of the spin precession and oscillation. The
parameter $T_0$ is the spin independent transmission amplitude and
$T_S$ is the spin dependent transmission originating from exchange
interaction between conduction electrons and localized
spin\cite{Teber10}.

The corresponding equation of motion reads
\begin{equation}
\frac{\partial {\mathbf S}}{\partial t}=-\gamma {\mathbf S}\times {\mathbf h}_{eff}
\label{spin-dyn}
\end{equation}
where $\gamma$ is the gyromagnetic ratio and ${\mathbf h}_{eff}$ is
the effective time dependent magnetic field including the applied
field as well as other contributions such as crystal anisotropy and
demagnetization fields. To produce a nutational motion the effective
magnetic field should contain two terms, a static magnetic field
along $z$ axis and a rotating transverse $rf$ field:
\begin{equation}
\mathbf{h}_{eff}(t)=\left(-h_0\sin\omega t\sin\Omega t,
h_0\sin\omega t\cos\Omega t, h_z\right).
\end{equation}
The solution of (\ref{spin-dyn}) is given by,
\begin{equation}
\label{spin-smm}
\nonumber\mathbf{S}(t)=S\left(\sin\theta(t)\cos\varphi(t),\sin\theta(t)\sin\varphi
(t),\cos\theta(t)\right)\,
\end{equation}
where $\varphi$ and $\theta$ are respectively the azimuthal and
polar angles of the molecular nanomagnet's spin in the spherical
coordinate. They are given by the following relations
\begin{equation}\label{theta&phi}
\varphi(t)=\Omega t,\;\;\theta(t)=\theta_0-\vartheta\cos\omega t.
\end{equation}
These relations describe the nutational motion of molecular
nanomagnet. It is a combination of precession around $z$ axis with
precession frequency $\Omega=\gamma h_z$ and tilt angle (angle
between the spin and $z$ axis) oscillation about $\theta_0$ with
amplitude $\vartheta$ and frequency $\omega=\gamma h_0/\vartheta$
(see Fig.(\ref{system-model})). There are many studies in the
literature which have focused on the properties of the system
emerging from the interplay of transported particles and localized
spin precession. However non of them investigated the effects of
localized spin nutation. In this work we study the simultaneous
effects of spin precession and spin oscillation ({\it spin
nutation}) on the transported charge carriers. In a very recent
work, we have shown that the spin nutation could be served as the
pumping parameters in such system and pump an $ac$ Josephson current
through the junction \cite{AAA12}. In the following we investigate
the effects of spin nutation on the Josephson current at the
presence of a bias voltage.

\section{Josephson current}\label{current}

In the tunneling limit and at the presence of a bias voltage $V$,
the Josephson current at lead $\alpha$ is given by
\begin{equation}\label{Josephson-current-operator}
I_{\alpha}^{J}(t)=
-e\int_{-\infty}^{t}dt'\left(e^{-i\frac{eV}{\hbar}(t+t')}\left\langle\left[A_{\alpha}(t),
A_{\alpha}(t')\right]\right\rangle+h.c.\right)\ ,
\end{equation}
where the operator $A_{\alpha}(t)$, is
\begin{equation}\label{a-operator}
\nonumber A_{\alpha}(t)=\sum_{k,k',\sigma\sigma'}c_{\alpha',k,\sigma}^{\dagger}(t)
T_{\sigma\sigma'}(t)c_{\alpha,k',\sigma'}(t)\ .
\end{equation}
In the above equation $\alpha=L(R)$ and $\alpha'=R(L)$. Let us
define the following retarded potential
\begin{equation}
\label{X-definition} X^{\sigma\sigma'}_{\rho\rho'}(t-t')=
-i\Theta(t-t')\left\langle\left[a_{k,k'}^{\sigma\sigma'}(t),a_{p,p'}
^{\rho\rho'}(t')\right]\right\rangle\ ,
\end{equation}
where we have defined
$a_{k,k'}^{\sigma\sigma'}(t)=c_{\alpha',k,\sigma}^{\dagger}(t)
c_{\alpha,k',\sigma'}(t)$. The retarded potential defined in Eq.
(\ref{X-definition}) contains both triplet and singlet correlations.

In the presence of a spin active junction including spin-flip
processes between a spin-singlet superconductor and a ferromagnet,
the singlet correlation penetrating into the magnetic region
converts to the triplet correlations. These induced triplet
correlations have the same magnitude as the singlet correlation at
the interface, but they survive over a long range despite of the
singlet correlation. As a result a nonzero Josephson current flows
through the strong ferromagnets (such as half metals)
\cite{Bergeret05,Braude07,Eschrig08}. The triplet correlations also
penetrate into the superconductors however their amplitudes are
small in comparison to the bulk singlet component
\cite{Halterman07}. The triplet correlations are also generated by
magnetization dynamics such as spin precession and magnonic
excitations. Interplay of the magnetization dynamics and transported
carriers results the conversion of the spin-singlet to spin-triplet
correlations which is accompanied by the absorbtion/emission of
magnons \cite{Takahashi07,Houzet08}. In this work, the presence of
the tunnel barriers indeed allows us to ignore the effects of the
triplet correlations induced in the superconducting leads and just
retain the singlet to triplet conversion process \cite{Takahashi07}.
Thus, the retarded potential (\ref{X-definition}) simplify as
$X^{\sigma\sigma'}_{\rho\rho'}(t-t')=\sigma\sigma'\delta_{\sigma,-\rho}
\delta_{\sigma',-\rho'}X_{ret}(t-t')$, where $\sigma,\sigma'=\pm 1$.
The associated Matsubara potential reads
\begin{equation}\label{X-Matsobara}
\mathcal{X}(i\omega_n)=\frac{1}{\beta}\sum_{iq}\mathcal{F}_R^{\dagger}(k,iq)
\mathcal{F}_L(k',iq-i\omega_n)\ ,
\end{equation}
where $\mathcal{F}_{\alpha}(k,iq)$ is the anomalous Green's function
\cite{Mahan90}. By analytical continuation and at the zero
temperature the real part of the retarded potential is obtained as
follows
\begin{equation}\label{Realpart-X}
\Re\left(\sum_{k,k'}X_{ret}(x)\right)=\left\{
\begin{array}{cl}
\pi N_LN_R\Delta K(x)\;& x<1
\\
\pi N_LN_R\frac{\Delta}{x}K(\frac{1}{x})\;& x>1
\end{array}\right.
\end{equation}
where $K(x)$ is the first kind of complete elliptic integral and
$N_{L,R}$ are the normal state density of states in the left and
right leads. Implementing the real part of the retarded potential
and defining
$\mathcal{J}(x)=\Re\left(\sum_{k,k'}X_{ret}(x)\right)/\pi N_LN_R$,
we find the scaled Josephson current $i^J=I^J/\pi e N_LN_R$ as
\begin{eqnarray}\label{Josephson-current}
\nonumber i^J(t)=&J_0\sin(\omega_{J}t+\chi)+\\
&\vartheta[J_{+}\sin(\omega_{+}t+\chi)+J_{-}\sin(\omega_{-}t+\chi)].
\end{eqnarray}
$\chi=\chi_R-\chi_L$ is the phase difference between superconducting
leads, $\omega_J=2eV/\hbar$ is the Josephson frequency and
$\omega_{\pm}=\omega_J\pm\omega$ represent the current oscillations
due to molecular nanomagnet spin nutation. In the above calculations
we have considered $\vartheta/\theta_0\ll 1$. This approximation is
fulfilled by the practical conditions of the system and dose not
make any restriction on it. The coefficients $J_0$ and $J_{\pm}$ are
given by the following relations
\numparts
\begin{eqnarray}
J_0=&4(T_{||}^2-T_0^2)\mathcal{J}\left(\frac{eV}{2\Delta}\right)
+2T_{\bot}^2[\mathcal{J}\left(\frac{eV+\hbar\Omega}{2\Delta}\right)+
\mathcal{J}\left(\frac{eV-\hbar\Omega}{2\Delta}\right)],
\\\label{Coefficients-Josephson-current}
J_{\pm}=&T_{||}T_{\bot}\bigg[2\mathcal{J}\left(\frac{eV}{2\Delta}\right)
+2\mathcal{J}\left(\frac{eV\pm\hbar\omega}{2\Delta}\right)
-\mathcal{J}\left(\frac{eV+\hbar\Omega}{2\Delta}\right)\\
\nonumber&-\mathcal{J}\left(\frac{eV-\hbar\Omega}{2\Delta}\right)
-\mathcal{J}\left(\frac{eV+\hbar\Omega\pm\hbar\omega}{2\Delta}\right)
-\mathcal{J}\left(\frac{eV-\hbar\Omega\pm\hbar\omega}{2\Delta}\right)\bigg]
,
\end{eqnarray}
\endnumparts
where $T_{||}=T_S\cos\theta_0$ and $T_{\bot}=T_S\sin\theta_0$ are
spin conserving and spin-flip transmission amplitudes, respectively.
The first term of Eq. (\ref{Josephson-current}) is the usual $ac$
Josephson current with oscillation frequency $\omega_J$ and the last
two terms are emerged due to the molecular nanomagnet spin nutation.

\section{Results and Discussions}\label{results}

We have obtained the Josephson current through the SC$|$molecular
nanomagnet$|$SC junction by considering the spin nutation and
applied bias voltage effects. In the absence of the tilt angle
oscillations ($\vartheta=0$ or equally $h_0=0$) the two last terms
in Eq.(\ref{Josephson-current}) vanish and the Josephson current is
$i^J(t)=J_0\sin(\omega_{J}t+\chi)$. This is nothing but the
conventional $ac$ Josephson current. The coefficient $J_0$ is
nonzero whenever one of the spin dependent amplitude ($T_S$) or spin
independent amplitude ($T_0$) is non-zero. The molecular nanomagnet
spin nutation causes to appear two $\vartheta$-dependent oscillatory
terms in the Josephson current. This effect strongly depends on the
values of $T_{||}T_{\bot}=\frac{T_S^2}{2}\sin 2\theta_0$, the
multiply of the spin conserving and spin flip transmission
amplitudes. The coefficient $T_{||}T_{\bot}$ is nonzero whenever
$\theta_0\ne n\pi/2$ ($n$ is an integer). Indeed the spin nutation
produces different potential energies for two electron spin
directions in the leads and the tunneling of the quasi-particles is
mediated by a spin flip process. These tunneling processes which are
accompanied by absorbtion/emission of the quantum of oscillations
result to appear the exotic additional oscillatory parts in the
Josephson current. These additional terms are explained by
implementing the Andreev levels picture. The Andreev levels are
sharp states in the superconducting gap produced by the constructive
Andreev reflections from the two superconducting interfaces. In
Refs. \cite{Holmqvist11,Holmqvist12} it has been shown that when the
nanomagnet undergoes a precession due to an external magnetic filed
the Andreev levels are affected by the dynamics of the nanomagnet's
spin. The resultant Andreev levels depend on the tilt angle and
precession frequency. In the adiabatic regime, when the title angle
oscillation is slow in comparison to the characteristic time of the
quasiparticles propagation, the Andreev levels dynamics will follow
the tilt angle oscillation. Thus, this oscillation modulates the
Josephson current in addition to the modulation due to the non-zero
bias voltage. In the tunneling regime where the {\it ac} Josephson
current is proportional to the $\sin(\omega_J t+\chi)$, the
sinusoidal time dependence of the tilt angle cause to appear two
additional terms in the Josephson current with modulation
frequencies $\omega_J\pm\omega$.

Numerical computations are needed to obtain the energy of Andreev
levels in the Josephson junction with a precessing spin and
arbitrary transparency of the contacts. An analytical expression for
the energy of Andreev levels in this junction has been provided in
Ref. \cite{Holmqvist12}, except a term giving rise to the Zeemann
splitting which has not been given an explicit expression for it. To
give an illustration for the above discussions let us calculate the
energy of Andreev levels by neglecting the tilt angle dependence of
this term and replacing it by an effective Zeemann splitting. In the
tunneling limit $T_0$ and $T_S$ are very small and the energy of
Andreev levels is given by
\begin{equation}\label{Andreev-levels}
\epsilon_{\pm}=-\Delta_0\sqrt{1+\Phi(T_0,T_s,\Omega,\theta,\chi)}\pm\frac{\hbar\Omega}{2\Delta},
\end{equation}
where
\begin{eqnarray}\label{Phi}
\Phi(T_0,T_s,\Omega,\theta,\chi)=&&-2T_0(1+\cos(2\theta))\sin^2(\chi/2)
\\&&-2T_s(1+\cos(2\theta))\cos^2(\chi/2)+
\left(\frac{\hbar\Omega}{2\Delta}\right)^2 .
\end{eqnarray}
The Josephson current at the zero temperature is given by
$d\epsilon_{-}/d\chi$. In the presence of an applied voltage we have
$\chi\rightarrow\chi+\omega_Jt $, and the Josephson current reads
\begin{equation}\label{Andreev-levels}
\mathcal{I}(t)=\frac{e\Delta}{\hbar}\frac{(T_0-T_s)}{\sqrt{1+\left(\frac{\hbar\Omega}{2\Delta}\right)^2}}
[1+\cos(2\theta(t))]\sin(\omega_J t+\chi)
\end{equation}
By expanding $\cos(2\theta)$ in terms of the small amplitude of tilt
angle oscillation $\vartheta$, the above expression reduces to
\begin{equation}\label{Andreev-levels}
\mathcal{I}(t)=\frac{e\Delta}{\hbar}\frac{(T_0-T_s)}{\sqrt{1+
\left(\frac{\hbar\Omega}{2\Delta}\right)^2}}[\cos^2\theta_0+\vartheta\cos(\omega
t)\sin\theta_0\cos\theta_0]\sin(\omega_J t+\chi)
\end{equation}
The term $\cos(\omega t)\sin(\omega_J
t+\chi)=(1/2)[\sin((\omega_J+\omega)t+\chi)+\sin((\omega_J-\omega)t+\chi)]$,
justify the appearance of the two terms in the {\it ac } Josephson
current which are proportional to $T_{||}T_\bot$ as indicated in Eq.
(\ref{Coefficients-Josephson-current}).

The nanomagnet's dynamics introduces two time dependent parameters
$\varphi(t)=\varphi_0+\Omega t$ and
$\theta(t)=\theta_0-\vartheta\cos(\omega t)$ specifying time
dependent direction of the nanomagnet's spin, in addition to the
applied voltage which causes the time dependent phase difference
across the junction, $\chi(t)=\chi +\omega_Jt$. The transport
properties of the Josephson junction depends on these time dependent
parameters. The combination of the Andreev reflection of
quasiparticles from the superconducing surfaces and their
interaction with the nanomagnet's spin dynamics lead to transfer of
the quasiparticles into sidebands with different energies given by
$\epsilon=\epsilon_0+\hbar\omega_J+m\hbar\Omega+n\hbar\omega$, where
$\epsilon_0$ is the energy of the quasiparticles in the absence of
these time dependencies and $m=n=0,\pm 1$. Scattering of the
quasiparticles to these sidebands leads to the appearance of
associated different terms in the coefficients $J_{\pm}$.

Because of the divergence of $K(x)$ at $x\sim 1$, when each of the
following nine values: $eV$, $eV\pm\hbar\Omega$, $eV\pm\hbar\omega$
and $eV\pm\hbar\Omega\pm\hbar\omega$ is equal to the superconducting
energy gap ($2\Delta$), the Josephson current diverges. These
divergences are due to the inadequacy of the lowest order
perturbation theory and they will disappear if we able to include
the higher orders of perturbation up to infinity, as it has been
discussed in Ref. \cite{Cuevas96}. More suitable approaches for
current calculation lead to singularities instead of divergences
\cite{Teber10,Holmqvist11}. These singularities are consequence of
the singular BCS density of states at the gap edges and are
signatures for contribution of an infinite number of extended states
in the current. These resonances emerge in the junction whenever the
absorbed/emitted energies of the transported quasi-particles, owing
to their interactions with the bias voltage and molecular
nanomagnet, are in the range of the superconducting energy gap
($2\Delta$). The typical superconducting gap is of the order of
$2\Delta\sim 1.0~ me{\rm V}$ and it can be smaller in an atomic
point contact. On the other hand, for a field $h\sim 200$ Gauss
(typical value of magnetic fields produced in a lab) the Larmor
frequency of the classical spin is about $560 ~{\rm MHz}$ and
exactly in the range of the relaxation time of the real magnetic
molecules. The energy associated with the precession energy is
$\hbar\Omega\sim\hbar\omega\sim 10^{-6}~e{\rm V}$ which is much
smaller than the typical superconducting gap. But, dependence on the
value of bias voltage provides the ability to simply tune the system
toward a resonance condition.

It is worthwhile to study the Josephson current when the molecular
nanomagnet oscillation frequency $\omega$ is the same as the bias
dependent frequency $\omega_J$. In this case $\hbar\omega=2eV$ and
the Josephson current is simplified as
\begin{eqnarray}
\nonumber i^J(t)=J_0\sin(\omega_{J}t+\chi)+\vartheta\left[J_-\sin\chi+
J_{+}\sin(2\omega_{J}t+\chi)\right]
\end{eqnarray}
As it is seen the current has two time dependent oscillatory and one
time independent non-oscillatory parts. It means that at the special
value of molecular nanomagnet oscillation frequency, the spin
nutation generates a time independent $dc$ Josephson current,
$\vartheta J_{-}\sin(\chi)$. In fact the effect of bias voltage is
partially washed out by spin nutation of the molecular nanomagnet
and a non-oscillatory part appear in the current. This is an
interesting result in which the spin nutation is able to convert the
$ac$ Josephson current to the $dc$ one.

\begin{figure}[h]
\centerline{\includegraphics[width=7cm]{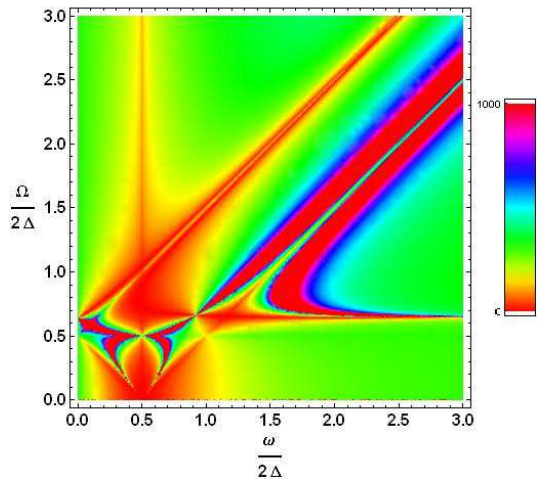}} \caption{(Color
online) Density plot of the ratio of the third and second terms of
the Josephson current ($|J_-/J_+|$) as a function of
$\frac{\omega}{2\Delta}$ and $\frac{\Omega}{2\Delta}$ for
$eV=\Delta$.} \label{amplitude-josephson}
\end{figure}

We have also studied the behavior of the Josephson current for
an arbitrary value of the bias voltage. In
Fig.(\ref{amplitude-josephson}) we have shown the density plot of
$|J_-/J_+|$ in terms of $\Omega$ and $\omega$ for $eV=\Delta$.
In the regions where $J_-/J_+\sim0$, the term oscillating with the
frequency $\omega_+$ has a dominant contribution. However in the
regions where $J_-/J_+$ tends to the infinity, the term with
frequency of $\omega_-$ becomes important. In a fixed value of $\omega$
we have one or both of the second or third terms of the Josephson
current, by varying $\Omega$.
%
%

\section{Summary and conclusion}

In this paper we have obtained the Josephson current through the
junction Superconductor$|$ molecular nanomagnet$|$ superconductor at
the presence of the bias voltage $V$. We have calculated the
Josephson current through the junction by working in the tunneling
limit and employing Green's function technique. We have investigated
the effects of the spin nutation, simultaneous effects of spin
precession and spin oscillation, on the behavior of the Josephson
current. The spin nutation is generated by applying a time dependent
magnetic field to the molecular nanomagnet. Interplay of the spin
dynamics of molecular nanomagnet and the Andreev reflection of the
quasiparticles at the superconductor surfaces which is accompanied
by spin-flip process while exchanging energy with the nutating spin
causes the generation of two extra terms in the Josephson current
with frequencies $\omega_J\pm\omega$, greater and smaller than the
usual Josephson frequency by the tilt angle oscillation frequency.
Depending on the values of the bias voltage, the precession
frequency and the tilt angle oscillation frequency, some divergences
emerge in the Josephson current. Such behavior appears whenever a
resonance occurs at the junction. The resonance conditions arise
when the superconducting energy gap is equal to the absorbed/emitted
energy of the transported quasiparticles, due to their interactions
with applied bias voltage ($V$) and with the dynamics of the
molecular nanomagnet. We have also found that the spin nutation is
able to convert the $ac$ Josephson current to the $dc$ one. This
effect may be used in the single spin detection.

It has been shown that a spin current is generated in the
superconducting leads by spin nutation of the molecular nanomagnet
\cite{Teber10,Holmqvist11,Holmqvist12,Stadler13}. The spin current
induces a torque on the spin and changes its dynamics. This
back-action effect leads to a change in the parameters of the
nutational motion {\it i.e.} precession frequency and tilt angle
oscillation frequency \cite{Holmqvist12,Zhu04}. Thus we can conclude
that it does not change our main results. Incorporating a Josephson
junction in a SQUID loop to control the phase difference of the
junction makes it possible to measure the current phase relation.
Recent advances in fabrication of molecular Josephson junctions has
made it possible to measure the current phase relation in an atomic
point contact \cite{Rocca07}. Tuning the bias voltage as
$2eV=\hbar\omega$ causes to appear a time independent term in the
Josephson current, whereas the two other terms oscillate with
frequency $\omega_J$. Taking the time average eliminates the effects
of the oscillatory terms. Changing the magnetic flux penetrating
into the loop changes the phase difference of the junction as well
as the coefficient of the time independent term $J_-$ without
altering the value of $\omega$. Therefore, measuring the time
average of the Josephson current will directly reveal the existence
of this time independent term.

We have considered an isolated molecular nanomagnet in our
calculations. A real molecular spin ${\mathbf S}$ interacts with its
environment \cite{Gatt07}. Thus, it's properties are different from
those of the isolated spin considered in our model. The interaction
of the spin with its environment such as exchange field, magnetic
anisotropy and ..., can be considered by including a small Gilbert
damping constant in the calculations \cite{Takahashi07}. In this
case, the dynamics of a single spin is given by the
Landau-Lifshits-Gilbert equation:
\begin{equation}
\frac{\partial {\mathbf S}}{\partial t}=-\gamma {\mathbf S}\times
{\mathbf h}_{eff} +\alpha{\mathbf S}\times\frac{\partial {\mathbf
S}}{\partial t} \label{spin-dyn2}
\end{equation}
where $\alpha$ is the Gilbert damping constant. At a finite
temperature the spin dynamics of a realistic nanomagnet is governed
by the stochastic Landau-Lifshits (SLL) equation \cite{Fesenko06}
\begin{equation}
\frac{\partial{\mathbf S}}{\partial t}=-\gamma{\mathbf S}\times
[{\mathbf h}_{eff}(t)+{\mathbf b}(t)]- \alpha[{\mathbf
S}\times[{\mathbf S}\times [{\mathbf h}_{eff}(t)+{\mathbf b}(t)]]],
\end{equation}
where ${\mathbf b}(t)$ is a stochastic field including the nonzero
temperature effects with the following statistical properties
\begin{equation}
\langle {\mathbf b}_{\alpha}(t)\rangle=0;~~~~\langle {\mathbf
b}_{\alpha}(t){\mathbf
b}_{\beta}(t')\rangle=2D\delta_{\alpha\beta}\delta(t-t'),
\end{equation}
where $\alpha$ and $\beta$ are the cartesian coordinates of the
field and $D$ is the strength of the thermal fluctuations. This
equation leads to a reduction of the size of spin caused by excited
state during the precession or reversal of the magnetization owing
to weak coupling with a thermal bath. We will consider the effect of
the temperature in the subsequent works.


\ack{J. Abouie and B. Abdollahipour acknowledge Shahrood University
of Technology where the initial parts of this work was done.}

\section*{References}

\end{document}